\documentclass[prb,noshowpacs,noshowkeys]{revtex4}
\pdfoutput=1
\usepackage{latexsym}
\usepackage{amsmath}
\usepackage{soul,xcolor}
\usepackage{times}
\usepackage{amssymb}
\usepackage{fancyheadings}
\usepackage[T1]{fontenc}
\usepackage[utf8]{inputenc}
\usepackage{url}
\usepackage{hyperref}
\usepackage{color}
\usepackage{graphicx}
 \usepackage{epsfig}
\usepackage{mathptmx}
\usepackage{bm}
\usepackage{array}
\usepackage{algorithm}
\usepackage{algorithmic}
\usepackage{url}
\usepackage{hyperref}
\usepackage{color}
\usepackage{soul,xcolor}
\setstcolor{red}

\usepackage{lscape}
\usepackage{draftcopy}

\begin{document}
\title{Multiscale phase oscillations induced by cluster
  synchronisation in human connectome core network}
\author{ Bosiljka Tadi\'c$^{1,2,3}$, Marija Mitrovi\'c Dankulov$^2$, Roderick Melnik$^{4,5}$}
\affiliation{$^1$Department of Theoretical Physics,  Jo\v zef Stefan Institute,
Jamova 39, Ljubljana, Slovenia}
\affiliation{$^2$Intitute of Physics, Pregrevica 118, 11080 Belgrade, Serbia}
\affiliation{$^3$Complexity Science Hub Vienna,  Metternichgasse 8,
  1030 Vienna, Austria}
\affiliation{$^4$MS2Discovery Interdisciplinary Research Institute;  M3AI Laboratory and Department of Mathematics, Wilfrid Laurier University, Waterloo, ON, Canada}
\affiliation{$^5$BCAM - Basque Center for Applied Mathematics; Alameda de Mazarredo 14, E-48009 Bilbao Spain}

\vspace*{3mm}
\date{\today}

\begin{abstract}
Brain imaging data mapping onto human connectome networks enables the investigation of global brain dynamics, where the brain hubs play an essential role in transferring activity between
different brain parts.  At this scale, the synchronisation processes are increasingly investigated as one of the key mechanisms revealing many aspects of brain functional coherence in healthy brains and revealing deviations due to various brain disorders.   For the human connectome core network, consisting of the eight brain hubs and the higher-order structure attached to them, previous simulations of Kuramoto phase oscillators at network nodes indicate instability of the global order parameter for a range of positive coupling strengths. 
In this work, we investigate the multiscale oscillations of the global
phase order parameter and show that they are connected with the
cluster synchronisation processes occurring in this range of couplings
below the master stability threshold. 
We use the spectral graph analysis and eigenvector localisation methodology, where the clusters of nodes playing a similar role in the synchronisation processes have a small mutual distance in the eigenvector space. 
We determine three significant clusters of brain regions and show the
position of hubs in them.
With the parallel analysis of the weighted core network and its binary
version, we demonstrate the primary role of the network's topology in the formation of synchronised clusters.  Meanwhile, the wights of edges contribute to the hub's synchronisation with the surrounding cluster, stabilise the order parameter variations and reduce the multifractal spectrum.
\end{abstract}
\maketitle

\section{Introduction\label{sec:intro}}

\textit{The Human connectome (HC)} structure representing the large-scale brain
networks is nowadays readily available from the brain-imaging data
\cite{HCP_Neuroimage} mapped to network structures. The  
nodes are identified as the anatomical brain regions  and the
edges represent neuron bundles connecting them.   
Integrated complex mapping procedures including the 
parcellation, tractography, and graph construction, are now
available subject to various selection of parameters; see,  for example, the Budapest connectome server
\cite{Budapest_HCserver3.0,Hung2} for details.
Advances in the knowledge of HC structure allowed investigation	of global
brain dynamics \cite{Brain_Hubs-dynamicsReview2018,SOC-Brain_meta2020,SOC_Brain_Gros2021}. In different
processes associated with the HC , the brain
hubs \cite{Brain_Hubs-development2019} as the best connected central
nodes play an essential role in transferring the activity between different
brain parts and the formation of co-activated groups of regions, bridging  between
anatomical and functional features \cite{ClSynch_SciRep2023anatom-funct}.  
It has been recognised how the  hub's fragility
\cite{Brain_Hubs_fragility2018} and 
 ``disturbances of their structural and functional connectivity profile are linked to
neuropathology'' \cite{Brain_Hubs-diseases2012} . Hence, the current
research efforts focuses on hubs-supported pathways in the brain function ; see, for
example,  studies in ref.\cite{HC-hubs-disorder2024} stating that  ``Understanding the concept of network hubs and their role in brain disease is now rapidly becoming important for clinical neurology''.

Based on identified brain networks,  the higher-order structures
 described by simplicial complexes, have been determined  in the
 consensus human connectomes \cite{Brain_weSciRep2019}. See also
 homological scaffold study in 
 \cite{Brain_Homology-Petri2014}.
The higher-order cortical connectivity is generally investigated as enabling reliable and efficient neural coding \cite{HOC_cortex-functions2024}. Higher-order structures are increasingly considered in 
the context of human brain functionality and, in particular, the
appearance and classification  of various brain disorders \cite{HOC_BrainDisorders2024,SOC-Brain_meta2020}.
In the human connectome network, the brain hubs appear mutually connected, representing a rich-club structure\cite{Brain_Hubs-richclub2011}. Moreover, the eight hubs hold the leading higher-order structures reaching different brain parts, thus representing the \textit{core of the human connectome} structure. In ref.\ \cite{we-Brain_SciRep2020}, the HC core network was determined by identifying simplexes of all orders attached to (at least one of)  the brain hubs. See detailed description below in Sec.\ \ref{sec:netsynch}.

\textit{Synchronization on human connectome} has been studied  
as one of the key processes of functional properties at the whole-brain scale, revealing the degree of collective dynamic behaviour in the interplay with the complex connectome's topology \cite{ClSynch_HConnectomeneuroimage2021}.
In this context, the use of a simple Kuramoto model with the oscillators
associated with the network nodes and influencing each other along the
weighted edges are widely accepted both in the theoretical consideration
as well as the empirical data evaluation
\cite{ClSynch_anaysisSciRep2020It,Synch_Kuramoto2021Neuroscience}.
In computational neuroscience, it has been recognised that the occurrence of synchronised clusters\cite{ClSynch_anaysisSciRep2020It} rather than the whole connectome synchrony corresponds to natural dynamical states in healthy brain functions supporting the emotion, cognition and behaviours; in contrast, the alternation of the functional connectivity and related patterns of partial synchrony
can be connected to specific types of brain disorders, neurologic and
psychiatric processes  \cite{Synch_Kuramoto2021Neuroscience}.

\textit{Cluster synchronisation in complex networks} has been
investigated in relation to the functional properties of networks and
their applications in neurology and other science disciplines\cite{ClSynch_chaoticReview2024,ClSynch_NeuroNLMech2014,ClSynch_anaysisSciRep2020It,ClSynch_vibrstab2023,ClSynch_methodFrontiers2023}.
Theoretically, in transition to a fully synchronised state, different clusters consisting of roughly synchronised nodes have different evolution patterns when the global coupling strength is below the master stability point; above this point, mutual synchronisation among different clusters occurs, leading to the complete network synchrony; see \cite{ClSynch_theoryBoccaletti} and references
there for detailed consideration.
Methodologically, clusters of nearly synchronised nodes are determined
using the spectral analysis of the Laplacian operator associated with
the underlying network structure \cite{ClSynch_NeuroNLMech2014}. The
methodology is based on the localisation of the eigenvectors of the sequence of smallest nonzero eigenvalues of the Laplacian, utilising the idea that nodes at ``symmetrical'' positions in the network, i.e., having a
vanishing distance in the eigenvectors space, are playing a similar
role in the synchronisation processes. See Section\ \ref{sec:clusters} for a more detailed description.

Building on the previous results in ref.\ \cite{BT_MCsynch2022}, we study the cluster synchronisation properties of the human connectome core network and their impact on the modulation of temporal dynamics of the global synchronisation order parameter.  
Using the methodology of the Laplacian spectral analysis, we
demonstrate the occurrence of cluster synchronisation in a range of
positive couplings associated with the observed order parameter
instability, and determine three leading clusters around the
brain hubs. 
A parallel analysis of the core network with the natural weights and its binary version reveals the primary role of topology
and the stabilising impact of the weight of the edges. These
cluster-synchronisation patterns dynamically impact the
multiscale fluctuations of the global order parameter, which we
study using the appropriate multifractal analysis. 

The paper is organised as follows. In Section\
\ref{sec:netsynch}, we present details regarding the
network structure, its spectral analysis, and the results of synchronisation
dynamics from \cite{BT_MCsynch2022} that are relevant to this
work. Furthermore, in Section\ \ref{sec:clusters}, we apply the
cluster-synchronisation methodology and determine the major cluster
both in the weighted HC core network and its binary version. Section\ \ref{sec:mfr} presents the multifractal analysis of the temporal
fluctuations of the order parameter inside the instability
region of both weighted and binary network versions.Final Section\
\ref{sec:conclusions} briefly summarises and discusses the results.

\section{Human connectome core network and its synchronisation
  properties\label{sec:netsynch}}
In this section, we give a more precise definition of the human
connectome core network and present its structural properties with the results of its spectral analysis.
Furthermore, as stated in the Introduction, we consider its synchronisation properties to elucidate the role of hubs in the appearance of clusters and their dynamic consequences. We show the results of the phase synchronisation of Kuramoto oscillators on this network \cite{BT_MCsynch2022} that motivate our study of cluster synchronisation. 

\subsubsection{The human connectome core network} 
\begin{figure*}[htb]
\begin{tabular}{cc} 
\resizebox{24pc}{!}{\includegraphics{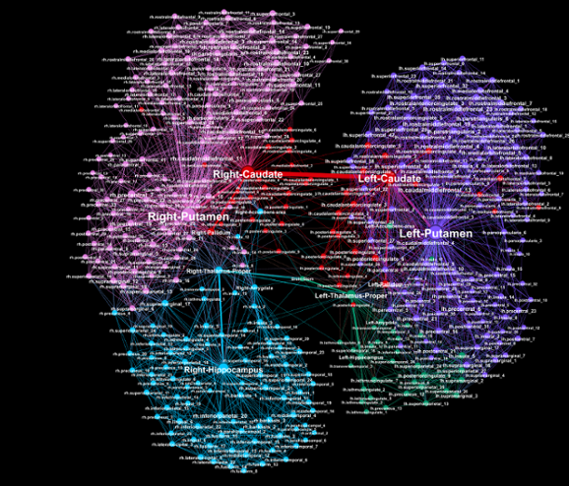}}
\end{tabular}
\caption{ The weighted undirected Human Connectome core network comprising the eight brain hubs and associated simplexes, with the weights on edges  extracted from the  consensus male connectome; data are from
  \cite{we-Brain_SciRep2020}. it contains $N=$418 nodes and $E=3835$ edges. The node's labels represent the anatomical brain area, the node's size is proportional to its degree, and colours indicate different topological communities. 
Graph visualisation using  \textit{gephi} \cite{Gephi-description}.}
\label{fig:netMc}
\end{figure*}

The core  connectome network is determined  \cite{we-Brain_SciRep2020} as a part of the human connectome
associated with the brain hubs
 \texttt{L. Caudate,
L. Putamen, L. Thalamus-Proper, L. Hyppocampus} in the Left and the corresponding Right hemisphere nodes. 
 Specifically, we use the HC core network shown
in Fig.\ \ref{fig:netMc}, which is  extracted from the consensus male
connectome determined in \cite{Brain_weSciRep2019}  
based on the brain imaging data from the Budapest human connectome 
server \cite{Budapest_HCserver3.0,Hung2}. 
The network consists of eight brain hubs and nodes (brain regions) that make simplexes of all orders associated with these hubs. Fig.\ \ref{fig:netMc} displays the structure of
connections associated with the brain hubs (big nodes) and weighted
edges, where the weights are determined according to the electrical
connectivity criteria as the number of fibres tracked between a pair
of brain regions normalised by the average fibre length between them
(see \cite{Brain_weSciRep2019} for details).

\begin{figure*}[htb]
\begin{tabular}{cc}
\resizebox{16pc}{!}{\includegraphics{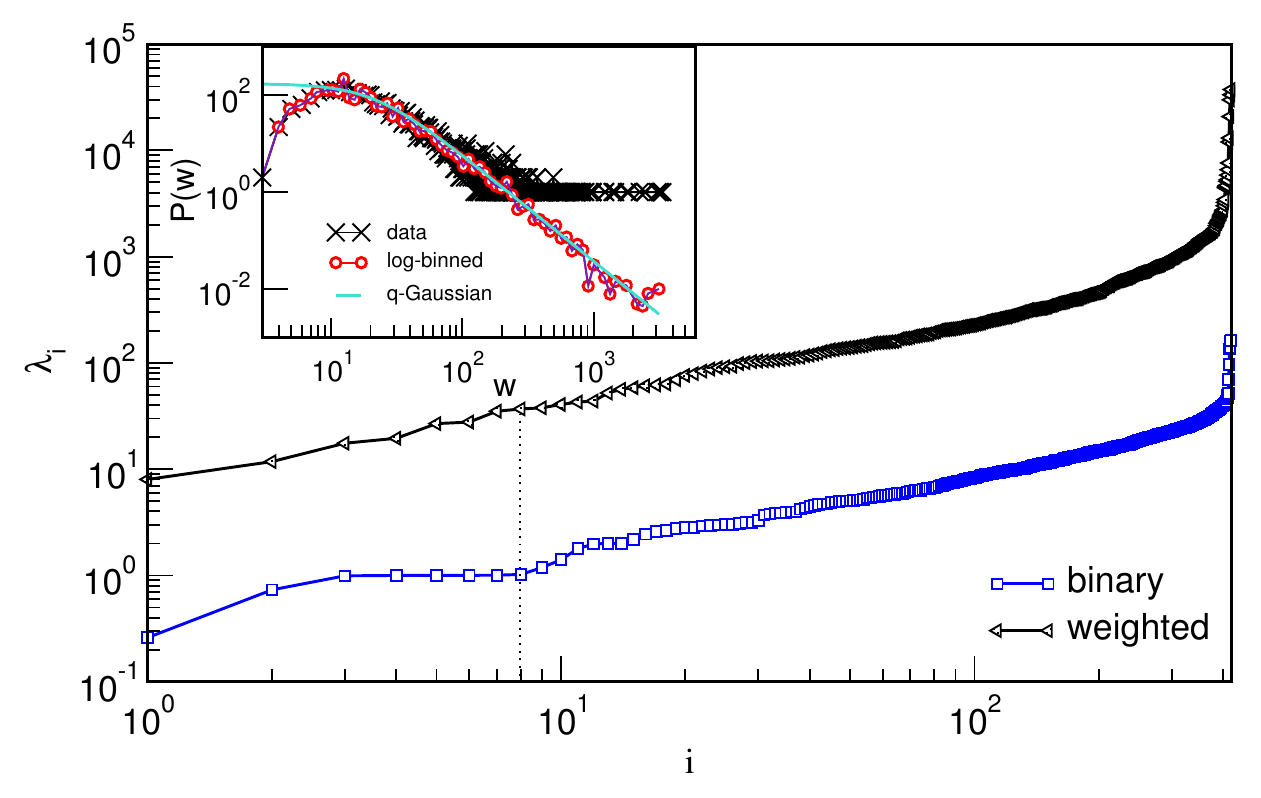}}&
\resizebox{14.2pc}{!}{\includegraphics{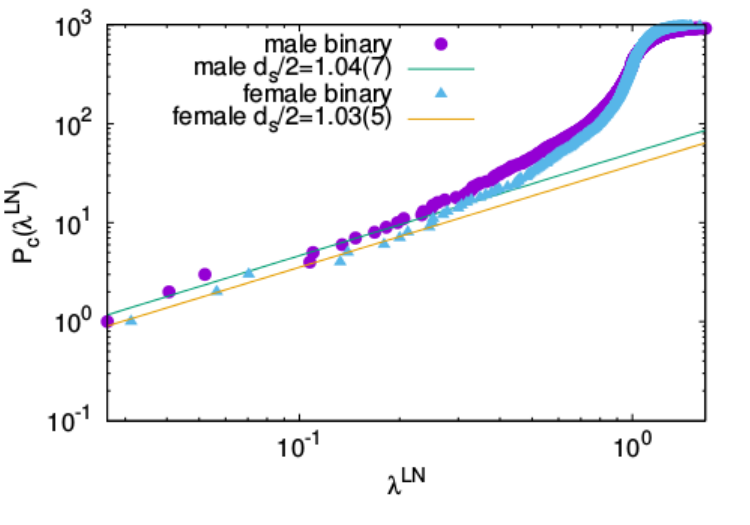}}\\
\end{tabular}
\caption{ Left panel shows the nonzero eigenvalues $\lambda_i$, $i=1,2,\cdots
  417$ in ranking order for the HC core network in  Fig.\ \ref{fig:netMc}. 
The distribution of weights of edges shown in the inset panel is fitted by 
$q-$Gaussian  with $q=1.9\pm0.11$; the data are from ref.\ \cite{we-Brain_SciRep2020}. The right panel shows the cumulative distribution $P_c(\lambda^{LN})$ of the 
 eigenvalues $\lambda^{LN}$ of the normalised Laplacian operator for
 the whole consensus human connectomes. The fitted regions for small
 $\lambda$ values determine the spectral dimension $d_s/2$.}
\label{fig:Pw-spectra}
\end{figure*}

The distribution of weights in this HC core network is shown in the inset of Fig.\ \ref{fig:Pw-spectra}, indicating a broad range of weights. Apart from the small weights, the data can be fitted by the $\tilde{q}$-Gaussian \cite{pavlos2014universality} distribution
$f(x)=a\left[1-(1-\tilde{q})\left({x/b}\right)^2\right]^{1/1-\tilde{q}}$, where we
find $\tilde{q}\lesssim 2$.
As shown in ref.\cite{we-Brain_SciRep2020}, all nodes and edges in
the male core connectome also exist in the consensus female connectome, but with slightly different weights on the edges. A detailed analysis of the higher-order connectivity using the simplicial complexes
representation of both female and male connectome and the
corresponding core networks can be found in refs.\ \cite{Brain_weSciRep2019,we-Brain_SciRep2020}.


\subsubsection{Eigenvalues spectrum of the human connectome 
  network} 
We focus on the spectral analysis of the Laplacian
operator associated with the human connectome networks (1-skeleton of
the simplicial complex), motivated by the theoretical background
connecting the eigenvalue spectrum of the
network structure with the diffusive dynamics on it \cite{Spectra_SD2003}.
Specifically, we determine the eigenvalues $\lambda _i$, $i=1,2 \cdots N$,  and the corresponding eigenvectors $v_{i}$
of the Laplacian of weighted and binary networks. For
convenience, we consider the positive-definite Laplace operator, whose
matrix elements $L_{ij}\equiv -\Delta_{ij}$ are given by
\begin{equation}
  L_{ij}=S_i\delta_{ij}-W_{ij}
  \label{eq:Lap}
\end{equation}
where $W_{ij}>0$ indicates the presence and the actual weight of the edge $(i,j)$ and
$S_i= \sum_{l=1}^{N} W_{il}$ is the node's strength. To
differentiate the impact of weights of the edges compared to the
global topology of the network in its synchronisation
features, we analyse in parallel the binary version of the HC core
network, in which $W_{ij}$ is replaced by the adjacency matrix
elements $A_{ij}=1$ for each $(i,j)$ pair where the edges exist, and zero
otherwise. Consequently, the $S_i\to k_i$ becomes the node's degree.
The main panel in Fig.\ref{fig:Pw-spectra} left shows the
eigenvalues of the Laplacian (\ref{eq:Lap}) for both the weighted core
network, Fig.\ \ref{fig:netMc}, and its binary version. The eigenvalues are
shown in the ranking order with $\lambda_0=0<\lambda_1 \leq \lambda_2
\leq \lambda_3 \leq \cdots \lambda_{N-1}$, where the equality applies
for the eigenvalue degeneracy. This figure shows the differences
between weighted and binary network spectra that affect their cluster
synchronisation. In particular, a range of small eigenvalues (up the
vertical dotted line) and their eigenvectors play a key role in the cluster synchronisation properties, as shown in Section\ \ref{sec:clusters}.
Before studying the cluster synchronisation in detail, we demonstrate
that the structure of the human connectome networks is suited for
partial synchronisation, featuring the spectral dimension $d_s\geq
2$. In the right panel of Fig.\ \ref{fig:Pw-spectra}, we show the
cumulative distribution of human connectome networks
(consensus male and female networks).  In this case, we use the
normalised \cite{Synchro_SAcliquesLaplacian2019} version of the
Laplacian (\ref{eq:Lap}), whose components are 
$L^{LN}_{ij}=\delta_{ij}-{{A_{ij}}\over{\sqrt{k_ik_j}}}$
    for the case of binary networks.  The cumulative distribution
    $P_c(\lambda)$  shows a scaling behavior $P_c(\lambda) \simeq
\lambda^{d_s/2}$ in the range of small nonzero eigenvalues. The
scaling exponent $d_s/2$ determines the network's spectral dimension
$d_s$, which signifies the impact of the network structure on the nature of stochastic dynamics on it; see, for example, references 
\cite{Spectra_noneqdynSciRep2018,Synch_spectraldimGinestra2019} and references there.  Here, we find that the spectral dimension of the human
connectome networks is close to $d_s=2.1$ within the numerical error
bars.  The spectral dimensions in this range are found in the networks
of self-assembled cliques of various dimensions in ref.\ \cite{Synchro_SAcliquesLaplacian2019}, suggesting a specific type of simplicial architecture with geometrically constrained aggregation.  According to the studies of network synchronisation, the structures with these spectral dimensions can not support thermodynamically stable synchronised phases \cite{Synch_spectraldimGinestra2019}.  Thus,
the occurrence of dynamical patterns with partial synchronisation is most probable.
Moreover, the components of the eigenvectors $v_{ik}$ of the smallest
nonzero eigenvalues of the Laplacian localise to the underlying
network communities and are efficiently utilised for the community detection \cite{mitrovic2009,Spectra-communitiesPRE2013}.  Recently, these spectral methods also gained attention in Data science and Machine learning literature \cite{Spectralpartitioning_Datasci2024,Spectraldensity_graphembedding2023}.
Fig.\ \ref{fig:netMc} shows that the considered human connectome core
network exhibits a community structure.  These communities appear to
affect the cluster synchronisation patterns, as shown in the next section.

\subsubsection{Synchronisation of phase oscillators on human connectome core
  networks} 
Simulations of synchronisation processes that we use in this work are done in ref.\ \cite{BT_MCsynch2022} considering  Kuramoto oscillators with the phase  $\theta_i$, $i=1,2,\cdots N$,  defined as an angle on a unit circle,  associated with the nodes of the core network in Fig.\
\ref{fig:netMc}, and interacting via the network's edges. The coupled equations govern the evolution of the node's phases 
\begin{equation}
    \dot{\theta_i} = \omega_i + \frac{K_1}{\sum_{l=1}^{N} W_{il}} \sum_{j=1}^{N} W_{ij} \sin{(\theta_j - \theta_i)}
    \label{eq:eomw}
\end{equation}
where  $K_1$ is the global coupling strength; the intrinsic frequency of the $i$-th oscillator $\omega_i$ is taken from Gaussian distribution and kept fixed during the simulation time. 
We note that the equation (\ref{eq:eomw}) takes into account the actual weights of the edges via $W_{ij}$ matrix elements and the normalisation factor, corresponding to the strength of the node $S_i\equiv \sum_{l=1}^{N} W_{il}$. As stated above, in the network's binary version, these factors are replaced by the adjacency matrix and the
node's degree, respectively. The network synchronisation properties
are studied in the literature by varying the global coupling strength $K_1$ from significant negative to large positive values. See also \cite{Synchro_we_PRE2021,Synchro_we_PRE2023} for synchronisation
of Kuramoto oscillators on simplicial complexes with different
architecture and simplex-embedded higher-order interactions. 

\begin{figure*}[htb]
\begin{tabular}{cc} 
\resizebox{24pc}{!}{\includegraphics{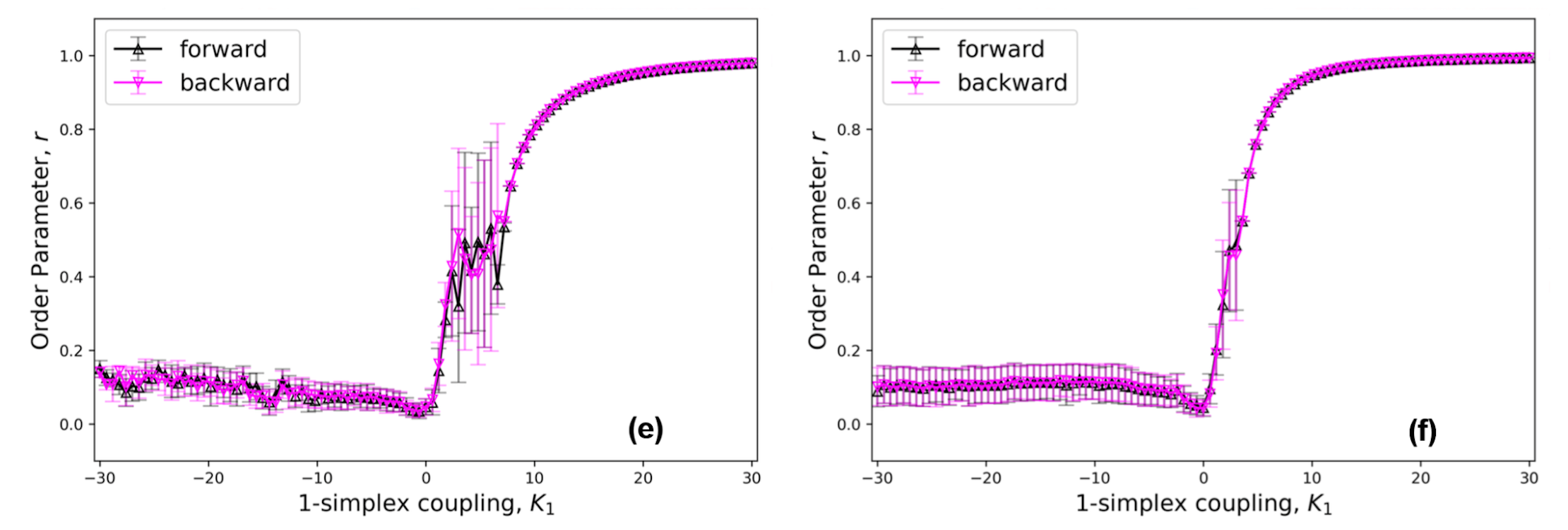}}\\
\end{tabular}
\caption{Phase synchronisation order parameter $r$  vs global coupling strength $K_1$ in HC core network. Data are part of fig2 in ref.\cite{BT_MCsynch2022}.}
\label{fig:OP2x}
\end{figure*}

A suitable measure of the collective synchronisation level is the Kuramoto order parameter $r$ defined as 
\begin{equation}
    r = \Biggl \langle \left|\frac{1}{N} \sum_{j=1}^{N} e^{i \theta_{j}}\right| \Biggr \rangle
    \label{eq:op}
  \end{equation}
  where $\langle .\rangle$ indicates the time average over a large number
  of simulation steps suitably chosen after a long transient
  interval. For the simulation details, see ref.\ \cite{BT_MCsynch2022}. We note that when a pattern of partial synchronisation occurs, the global
  order parameter  may have finite values $0 <r < 1$ between the
  completely desynchronized phases, where $r \to 0$, and full
  synchrony $r\to 1$, which can be reached for $K_1>0$ asymptotically when the global  coupling $K_1\to \infty$.
Fig.\ref{fig:OP2x} shows the results for the order parameter for the phase
synchronisation obtained in ref.\cite{BT_MCsynch2022} on
the above-described human connectome core network,
Fig.\ref{fig:netMc}, and its binary version. This figure shows that
both weighted and binary networks exhibit instability in the region
of positive couplings $K_1$, here denoted as 
$[K_1^-,K_1^+]$, where the (time and sample) averaged order parameter fluctuates. Such instability occurs for a range of small coupling strengths
$K_1>0$, before the full synchrony is reached asymptotically at
substantial $K_1$ values. Notably, this instability is more
pronounced in the case of binary structure than in the real weighted
network. Based on the temporal evolution of individual node's phases studied in \cite{BT_MCsynch2022}, we can identify the lower bound $K_1^-$ where the nontrivial grouping of the evolution paths first occurs. We then follow the cluster aggregation until three major clusters eventually emerge, which mutually coordinate their evolution above the upper boundary $K_1^+$. (The whole video of the simulations in \cite{BT_MCsynch2022}  is available online \url{https://youtu.be/E9nFrxuoplk} for the range $K_1\in [-30,30]$ in small $dK$ steps, 
where the starting state at $K_1\to K+dK$ value is achieved for the
preceding $K_1$, the procedure is known as ``tracking the attractor'' \cite{Synchro_we_PRE2021}).
Hence, we can identify these bounds  for the binary network as $[1.88, 7.2]$  within which we observe nonrandom fluctuations
of the order parameter that will be associated with the cluster
synchronisation, as shown below; meanwhile, the weighted edges lead to a more stable synchronisation pattern and shrinking
the instability region, estimated as   $[0.9,4.2]$ for the original weighted
network; see Section\ \ref{sec:clusters}.

\begin{figure*}[htb]
\begin{tabular}{cc} 
\resizebox{24pc}{!}{\includegraphics{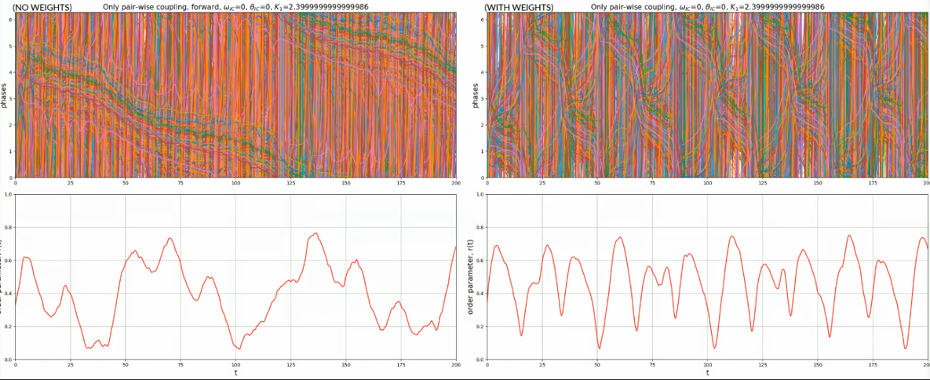}}\\
\end{tabular}
\caption{Time evolution of phases of each node in the instability region for binary (left) and weighted HC core network  (right);  lower
  panels show the corresponding order parameter $r$ vs time $t$ in steps of $0.01$; data are from Ref.\cite{BT_MCsynch2022}.}
\label{fig:Phases2x}
\end{figure*}
In Fig.\ \ref{fig:Phases2x}, the simulation data
\cite{BT_MCsynch2022} of the phase evolution of individual nodes are
shown to demonstrate the differences between the weighted network and its binary version; these data refer to the instability region, $K_1=2.399$ of both networks. Several groups of roughly synchronised nodes are visible in both cases, moving at different speeds. In the following section, we show the occurrence of synchronised clusters behind these temporal profiles. These evolution
patterns lead to different temporal fluctuations of the global synchronisation order parameter, eq.\ (\ref{eq:op}), as shown in the
lower panels of Fig.\ \ref{fig:Phases2x}.  Quantitative differences
between these modulated oscillatory patterns of the order parameter are revealed by the multifractal analysis in section\ \ref{sec:mfr}.

\section{Cluster synchronisation in HC-core
  network\label{sec:clusters}}
As stated above, to determine synchronised clusters, we use the
approach based on the  localisation of the eigenvectors of the
Laplacian operator (\ref{eq:Lap}) associated with the HC weighted core
network, and also for its binary version described above. The
corresponding eigenvalue spectra are shown in the left panel of Fig.\
\ref{fig:Pw-spectra} with the ordered eigenvalues $\lambda_0<\lambda_1\leq
\lambda_2...\leq \lambda_{N-1}$.
As described in the literature, see, for example, ref.\ \cite{ClSynch_methodFrontiers2023},  this
methodology resides on the idea of the master-stability 
threshold $g_c\equiv K_c\lambda_1$ for the global pairwise coupling $K_c$, above which the complete synchrony associated with the leading eigenvalue may emerge. Meanwhile, below $g_c$, the partial synchronisation can occur in terms of different clusters of nodes with
almost identical evolution associated
with the eigenvalue modes of $\lambda_0<\lambda_1\leq
\lambda_2 \cdots \leq \lambda_m$  orthogonal to the main synchronisation pattern \cite{ClSynch_methodFrontiers2023}.   (Note that we restrict to the pairwise coupling; recently, the master-stability equation was generalised to consider simplexes larger than network edges in ref. \cite{SC_MasterstabilityEq2022}.)

Assuming that the threshold $K_c=K_1^+$ associates with the upper
bound of the instability region discussed above, then the sequence of
modes corresponding to  the eigenvalues $\lambda_i$
is related to smaller values of the global coupling $K_{1i}$ such that
\begin{equation}
K_{1i} = K_c\lambda_1/\lambda_i  
\label{eq:Ki}
\end{equation}
where $ i=2,3,\cdots m$  until the lower bound is reached, i.e.,  $ K_{1i} \geq K_{1}^- $.  Thus, we determine the sequence of $m$ eigenvalues that correspond to
these unstable modes and their eigenvectors for both binary and
weighted HC core network.  
Nodes that have the same projections of eigenvectors and, thus, a small
distance in the space of eigenvectors corresponding to
these modes are considered to belong to the same cluster. More
precisely, for nodes
$i$ and $j$ we can define distance 
\begin{equation}
\delta e_{ij}=\sum^{m+1}_{k=1}|v_{ki}-v_{kj}|  < \theta
\label{eq:distance}
\end{equation}
where $v_{ki}$ and $v_{kj}$ are $i$ and $j$ components of the vector
$k$ that corresponds to $k$ unstable mode. Ideally, $\theta=0$;
however, due to numerical precision limitation, a finite threshold is
assumed, which we chose as $\theta=0.003$, according to ref.\ \cite{ClSynch_methodFrontiers2023}.\\
Given the differences in the eigenvalue spectra, cf.\ Fig.\ \ref{fig:Pw-spectra}, the number of unstable modes significantly differs in the weighted and binary networks. Precisely, we have $m=8$ different modes in the weighted network; meanwhile, the eigenvalues $3<i<8$ are degenerate and  only three first modes with different values
contribute in the case of the binary network; cf.\ table\ \ref{tab:EVs}. 
\begin{table}
\label{tab:EVs}
\begin{tabular}{ccc}
\hline
i& (binary) &weighted\\
1&0.25935471496502077 &8.00624257944399 \\
2& 0.7319926928674106 &11.766498485351391 \\
3&0.9904154802265737& 17.52885798614962\\
4&1.0000000000000195& 19.519470842538514\\
5&1.0000000000000229& 26.827478845152257\\
6&1.000511313383658& 27.712076731684096 \\
7&1.0047717952418098&35.09829539136315 \\
8& 1.020652012416054& 36.86219625823234 \\
\hline
\end{tabular}
\end{table}
With the equation (\ref{eq:Ki}) and the upper bound $K_1^+$ given above,
these eigenvalues lead to the sequence of global couplings
$K_{1i}=$2.551,1.885, 1.867 for $i=2,3,4$ for the binary network. 
The corresponding  values for the weighted network are
2.857, 1.918, 1.733, and so on; we have $K_{i=7}=0.958$ and two more values above the lower bound identified as $K_1^-=0.9$. 

\begin{figure*}[!htb]
\begin{tabular}{cccc} 
\resizebox{24pc}{!}{\includegraphics{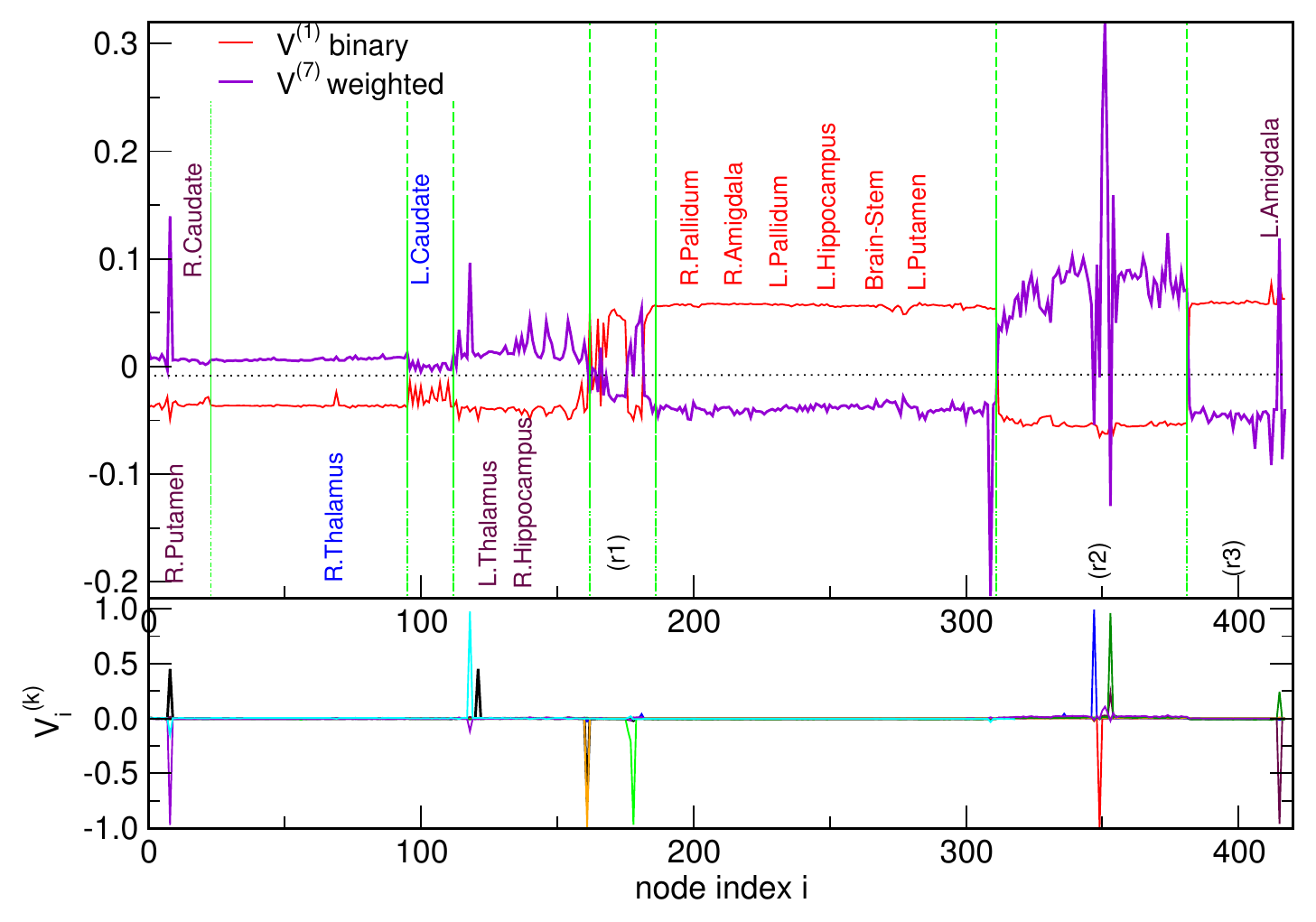}}\\
\end{tabular}
\caption{The Laplacian eigenvectors of leading
  eigenvalues for the binary and weighted network localised on several
  nodes (lower panel) and two delocalised eigenvectors indicated in
  the legend (upper panel). Positions of the brain hubs are indicated;
  see text for more. }
\label{fig:Vectors}
\end{figure*}

The eigenvectors corresponding to these eigenvalues are shown in Fig.\
\ref{fig:Vectors}. Notably, most of the eigenvectors in this region of
the spectra appear to localise on one or a couple of nodes; cf.\ Fig.\ \ref{fig:Vectors} lower panel. Meanwhile, the eigenvectors $v_1^{B}$ of the binary
and $v_7^{W}$ of the weighted network version are delocalised 
over four different network parts and the transition regions
between them. Apart from more pronounced fluctuations of the
eigenvector components in the case of $v_7^{W}$, the nodes and network regions where these eigenvectors have nonzero components are practically identical, suggesting the primary impact of topology in the eigenvector (de)localisation, cf.\ Fig.\ \ref{fig:Vectors}.

\begin{figure*}[!htb]
\begin{tabular}{cccc} 
\resizebox{16pc}{!}{\includegraphics{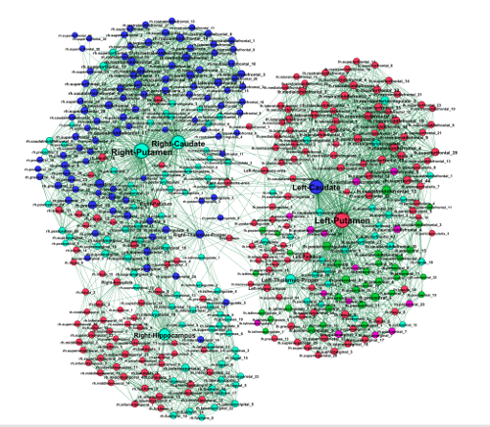}}&\resizebox{16pc}{!}{\includegraphics{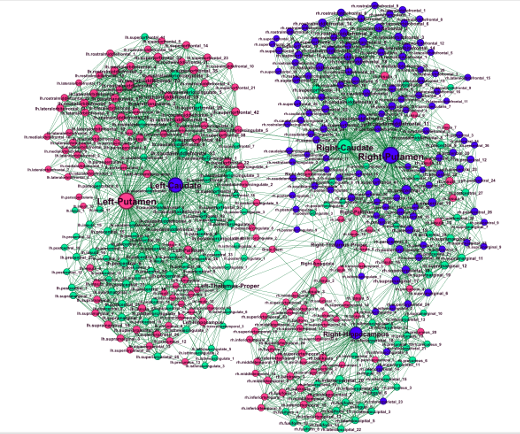}}\\
\end{tabular}
\caption{The three most significant clusters  (indicated by the blue, red and green
  colour of nodes) are identified across the instability region of
  the Human Connectome core network with the natural weights of the
  edges (right) and its binary version  (left). All remaining nodes
  are shown in magenta. The size of nodes and
  their names (brain regions) are provided, scaled with the node's degree. 
}
\label{fig:clusters_b4x}
\end{figure*}
Furthermore, we determine synchronised clusters using the
eq.(\ref{eq:distance}) with the above-mentioned number of contributing
eigenmodes in the weighted and binary network. Specifically, we find
three large clusters and a number of small ones consisting of a few
nodes and a certain number of nodes that do not belong to any of
the clusters. In Fig.\ \ref{fig:clusters_b4x}, we present the HC core
binary and weighted networks with the identity of nodes belonging to three leading clusters shown in red, green and blue colour. All smaller clusters and unsynchronised nodes are shown in magenta colour. We note that apart from some hubs, the nodes belonging to these clusters in the weighted and binary network coincide to a large extent, which agrees with the eigenvector localisation in Fig.\ \ref{fig:Vectors}. Moreover,
some hubs appear out of these clusters in the binary network,
e.g., R.Putamen and R.Hyppocampus; meanwhile, in the weighted network, they are synchronised with the surrounding blue and red clusters, respectively.

\section{Multiscale temporal fluctuations in the cluster-synchronisation region\label{sec:mfr}
}
The occurrence of cluster synchronisation manifests in the modulations
of the temporal evolution of the global order parameter $r$. We proceed
with the multifractal analysis of the order-parameter time series for
a coupling strength $K_1=2.399$ inside the instability region for the 
weighted and binary network;  see lower
panels of Fig.\ \ref{fig:Phases2x}. 

\begin{figure*}[htb]
\begin{tabular}{cc} 
\resizebox{24pc}{!}{\includegraphics{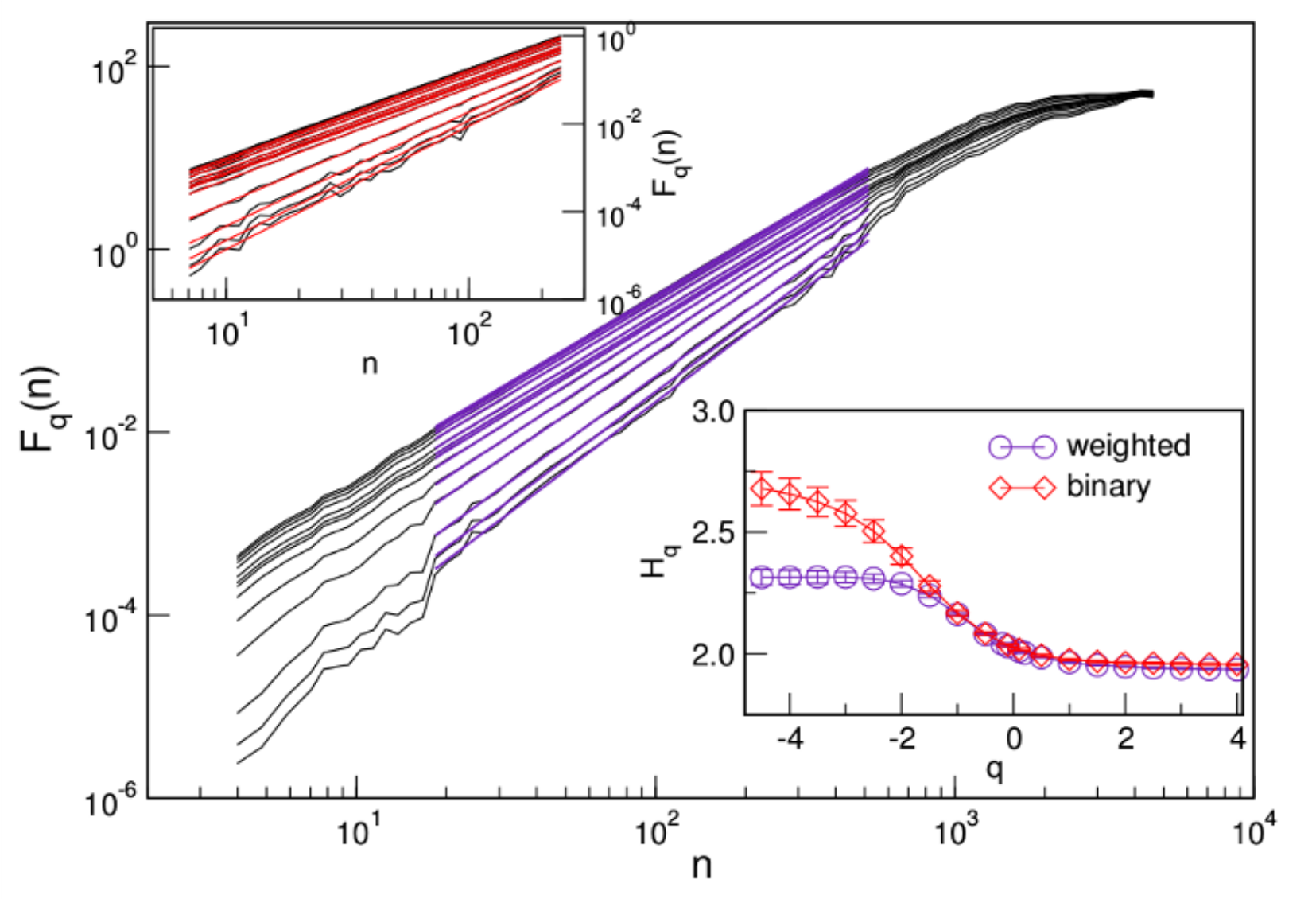}}
\end{tabular}
\caption{The fluctuation function $F_q(n)$ vs time interval $n$ for the order-parameter fluctuations in the instability region, $K_1=2.399$, shown in Fig.\ \ref{fig:Phases2x}. Thick lines indicate the fitted region. 
 The central panel shows the weighted core network's fluctuation function, while the binary network case is shown in the top inset. The corresponding generalised Hurst exponents $H_q$ vs $q$ are given in the bottom inset. }
\label{fig:Fq_w}
\end{figure*}

We use the multifractal analysis of time series \cite{MFRA-uspekhi2007,DMFRA2002,dmfra-sunspot2006,gao_JSTAT2009,dmfra_drozdz2015,MFRA_JSTATbhn2016} to determine the
nature of the OP fluctuations $r(t_k)$, for different datapoints
$t_k=1,2,\cdots T_{max}$; in the simulations \cite{BT_MCsynch2022}, $dt=0.01$ such
that the evolution time is $T_{max}=20000$. 
We then use these multifractal properties  to quantify the difference between the
weighted and binary networks inside the instability region.
The applied methodology utilises the self-similarity of the
generalised fluctuation function $F_q(n)$ against the varying time
interval $n$; see Fig.\ \ref{fig:Fq_w}.
Specifically, the fluctuation function is determined
\cite{DMFRA2002,MFRA_JSTATbhn2016} using the time-series profile $Y(i)
=\sum_{k=1}^i(r(t_k)-\langle r\rangle)$, which is then divided into non-overlapping segments of the length $n\in [2,T_{max}/4)$; the
process is done once starting from the
beginning of the time series and again starting from its end, thus
resulting in $2N_s=2Int(T_{max}/n)$ segments. 

For  each segment $\mu=1,2\cdots N_s$ the
local trend $y_\mu(i)$ is determined by using an interpolation procedure, and the standard deviation around that trend is defined as
 $F^2(\mu,n) = \frac{1}{n}\sum_{i=1}^n[Y((\mu-1)n+i)-y_\mu(i)]^2 $, 
and similarly $F^2(\mu,n) =
\frac{1}{n}\sum_{i=1}^n[Y(N-(\mu-N_s)n+i)-y_\mu(i)]^2$ for $\mu
=N_s+1,\cdots 2N_s$ in the opposite direction.  The $q$-th order
fluctuation function $F_q(n)$ is then obtained 
\begin{equation}
F_q(n)=\left\{(1/2N_s)\sum_{\mu=1}^{2N_s} \left[F^2(\mu,n)\right]^{q/2}\right\}^{1/q} \sim n^{H_q}  
\label{eq-FqHq}
\end{equation}
for varied segment length $n$ and its scale invariance investigated
for a range of the values $q\in \cal{R}$. 
For each $q$ value, the segments that appear to be scale-invariant
(seen as straight lines in a double-log plot) are fitted to determine
the corresponding scaling exponent $H_q$, defined in eq.\ (\ref{eq-FqHq}).  
The spectrum of different generalised Hurst exponents
expressed by the function $H_q$  vs $q$ characterise the
multifractal features of the time series, and we have $\Delta
H=H_q^{max}-H_q^{min} > 0$, indicating how different segments of the time series need to be amplified differently to become self-similar to the rest of the data points. Here, we vary $q\in [-4.5,+4.5]$ such that $H_q$ saturates in both limits. Thus, we examine the small-fluctuation segments, which are enhanced for the negative
$q$ values, and the large fluctuations segments that dominate the
fluctuation function for the positive $q$ values.

The results are given in Fig.\ \ref{fig:Fq_w}, depicting the differences
between the global phase fluctuations in the weighted and binary HC
core networks. In particular,   the fluctuation function $F_q(n)$ vs
time interval $n$ is shown with the corresponding scaling regions and
the resulting multifractal spectra $H_q$ vs $q$. Notably, at this
point of the instability region, we have 
multifractal spectra with $\Delta H_q^{B}>\Delta
H_q^{W}>0$. Specifically, the spectra differ significantly at $q<0$ side, corresponding to the enhanced small-scale fluctuations in the
binary network, compared to the weighted counterpart. On the other
side, the large fluctuations corresponding to $q>0$ side of the
spectra are of the same type and close to the fractional Brownian
motion in both networks.

\section{Discussion and Conclusions\label{sec:conclusions}}
Based on the human connectome core network, we presented a
comprehensive methodology for analysing the multiscale oscillations of
the global phase order parameter occurring in a range of positive couplings below the full synchrony threshold.
We have expanded the previous study \cite{BT_MCsynch2022} of
simulations of synchronisation processes by considering the
cluster synchronisation with the Laplacian eigenvectors localisation and multifractal analysis of temporal evolution of the synchronisation
order parameter. We demonstrated how these oscillations are linked to
cluster synchronisation processes across a range of positive coupling
strengths, all while remaining below the master stability
threshold. An important aspect of our approach is based on spectral
graph analysis of brain networks \cite{Brain_spectral2021,Brain_spectra2025NetwNeurosci} and eigenvector localization. It helps us locate clusters of nodes that play similar roles in synchronisation processes; these clusters are closely positioned within the eigenvector space. Consequently, we can identify the key clusters of brain regions and pinpoint the positions of hubs within these clusters.
Furthermore, given the complexity of the network architecture we are working with, we carried out a parallel analysis of both the weighted core network and its binary version. It highlights the crucial role that the network's topology plays in forming synchronised clusters.
Specifically, our results revealed that:
\begin{itemize}
\item Different synchronised clusters occur for a range
  of positive coupling strengths in the human connectome with the
  empirical weights and its binary version; this conclusion is
  supported by finding the connectome's spectral dimension is $d_s\eqsim 2.1$; 
\item Similarity of nodes identified in three major clusters surrounding brain hubs
  for weighted and binary network versions suggest the primary role of
  topology in the cluster synchronisation processes; 
\item Presence of these clusters causes multifractal fluctuations
  of the global phase order parameter, leading to different
  singularity spectra in the weighted and binary network version.
\end{itemize}

Several open questions remain for future study, in particular, given a complex higher-order structure of human connectome \cite{Brain_weSciRep2019,we-Brain_SciRep2020} and the current trend of research in the synchronisation processes on
higher-order networks \cite{Synch_HOCreview2024,HOCsoc_weEPJB2024} and their
applications in neurosciences
\cite{ClSynch_HConnectomeneuroimage2021,ClSynch_anaysisSciRep2020It,Synch_Kuramoto2021Neuroscience}. 
Particularly in contrast to interacting spin systems on simplicial
complexes, where triangle-embedded interaction appears as another
fundamental interaction inducing the collective (self-organised) state
\cite{HOCsoc_weEPJB2024}, 
the transition to a synchronised state is enabled solely by increased pairwise coupling strengths. Meanwhile, the higher-order interactions have an impact limited to the appearance of abrupt desynchronisation
\cite{HOC_Synchro_ArenasPRL2019} and the shape of the hysteresis loop, depending on the dimension of the oscillators \cite{HOC_Synch_Boccaletti2021} and size of simplexes in the simplicial complex, its architecture and the distribution of
internal frequencies
\cite{Synchro_we_PRE2021,Synchro_we_PRE2023}.
Therefore, a nontrivial impact of higher-order interactions on the cluster
synchronisation can be expected, in particular, in such complex
structure, such as human connectome; this question 
remains to be investigated. We note that a different type of partial
synchronisation exhibiting many small clusters due to frustration
effects was observed for a broad range of negative pairwise couplings
\cite{BT_MCsynch2022}. 
There are still open questions regarding the differences in cluster
synchronization between female and male connectomes, 
the resting state versus the active brain, and the recognition of
brain disorders.
We observe a certain degree of analogy in our approach to
order-parameter-based multifractality analysis when studying cluster
synchronization in human connectome core networks to spectral
methodologies for other localization and delocalization phenomena,
such as Anderson localization in complex systems. We also note that to
the best of our knowledge, the first proposal for treating
neuropathological conditions inspired by Anderson localization was
introduced in \cite{AndLoc_JRSIF2017}. The theoretical framework we have developed here opens new avenues for research in this direction.

In summary, we demonstrated how the topology around brain hubs critically shapes the global brain's dynamical coherence. In this context, cluster synchronisation plays a significant role in providing
mechanisms for partial synchrony associated with the human connectome topology and quantified by the multifractal features of the global phase order parameter.
In connection with empirical data, the study of these mechanisms can  manifest further differences in brain function when  processing
specific tasks or deviations related to brain disorder pathways.

\acknowledgements
B.T.  acknowledge the financial support from the Slovenian
Research Agency under the program P1-0044. RM thanks the NSERC and the
CRC Program for their support. MMD acknowledges funding provided by the Institute of Physics Belgrade, through the grant by the Ministry
of Education, Science, and Technological Development of the Republic of Serbia. Numerical simulations were run on the PARADOX-IV supercomputing facility at the Scientific Computing Laboratory, National Center of Excellence for the Study of Complex Systems, Institute of Physics Belgrade.

\bibliographystyle{unsrt}

\end{document}